\documentclass[conference]{IEEEtran}
\IEEEoverridecommandlockouts
\usepackage{cite}
\usepackage{amsmath,amssymb,amsfonts}
\usepackage{mathtools}
\usepackage{algorithm}
\usepackage{algpseudocode}   
\usepackage{graphicx}
\usepackage{textcomp}
\usepackage{xcolor}
\usepackage{stmaryrd}
\usepackage{bbm} 
\usepackage{booktabs}
\usepackage{hyperref}
\def\BibTeX{{\rm B\kern-.05em{\sc i\kern-.025em b}\kern-.08em
    T\kern-.1667em\lower.7ex\hbox{E}\kern-.125emX}}
\begin{document}

\title{Efficient Privacy-Preserving Recommendation on Sparse Data using Fully Homomorphic Encryption
\\

}

\author{\IEEEauthorblockN{Moontaha Nishat Chowdhury}
\IEEEauthorblockA{
\textit{Independent Researcher}\\
chowdhurymoontaha3@gmail.com}
\and
\IEEEauthorblockN{Andr\'{e} Bauer}
\IEEEauthorblockA{
\textit{Illinois Institute of Technology}\\
abauer7@iit.edu}
\and
\IEEEauthorblockN{Minxuan Zhou}
\IEEEauthorblockA{
\textit{Illinois Institute of Technology}\\
mzhou26@iit.edu}

}

\newif\ifcomments
\commentstrue

\newcommand{\minxuan}[1]{\ifcomments\textcolor{blue}{\sf\bfseries    Minxuan: #1       }\else\fi}
\newcommand{\moontaha}[1]{\ifcomments\textcolor{red}{\sf\bfseries    Moontaha: #1       }\else\fi}

\maketitle

\begin{abstract}
In today's data-driven world, recommendation systems personalize user experiences across industries but rely on sensitive data, raising privacy concerns. Fully homomorphic encryption (FHE) can secure these systems, but a significant challenge in applying FHE to recommendation systems is efficiently handling the inherently large and sparse user-item rating matrices. FHE operations are computationally intensive, and naively processing various sparse matrices in recommendation systems would be prohibitively expensive. Additionally, the communication overhead between parties remains a critical concern in encrypted domains. We propose a novel approach combining Compressed Sparse Row (CSR) representation with FHE-based matrix factorization that efficiently handles matrix sparsity in the encrypted domain while minimizing communication costs. Our experimental results demonstrate high recommendation accuracy with encrypted data while achieving the lowest communication costs, effectively preserving user privacy.
\end{abstract}


\section{Introduction}
Recommendation systems are widely deployed to help various customers discover preferred products and content, such as online shopping or browsing streaming services. Collaborative filtering (CF) is a commonly used algorithm for recommendation systems \cite{b1}. It infers the user’s preference from the gathered history of other users and provides a recommendation for new items that are similar to the user’s preference. However, several studies have noted serious privacy concerns during CF. \par
The recommendation system may provide users’ private data to third parties for a profit without the users’ permission \cite{b2}. Narayanan and Shmatikov presented a method to identify users by comparing the anonymized public Netflix dataset to the Internet Movie Database (IMDb) \cite{b3}. Weinsberg et al. showed that user rating lists can compromise privacy by revealing sensitive information including users' age, gender, ethnic background, and political views \cite{b4}. To address these privacy issues, numerous works have proposed privacy-preserving CF algorithms \cite{b5,b6,b7, b8}, using anonymization, differential privacy, and cryptographic tools. However, anonymization is not sufficient to protect privacy \cite{b3,b5,b9}, and differential privacy causes a tradeoff between privacy and prediction accuracy. A promising alternative for privacy-preserving recommendation systems is applying cryptography protocols to protect the data in the encrypted domain. 
Recent works have achieved significant advancements in protecting user privacy in Matrix factorization, a core technique in CF, by utilizing various cryptography protocols\cite{b7, b23}. Specifically, Nikolaenko et al.\cite{b7} proposed a hybrid cryptographic approach using partially homomorphic encryption \cite{b17} and garbled circuits \cite{b18} to protect user ratings, though the item profile remains unencrypted and iterative computations are inefficient due to large circuit sizes and heavy communication overhead. Kim et al.\cite{b11} introduced a secure multi-party computation (MPC) protocol \cite{b34} involving users, a recommendation system (RS), and a crypto service provider (CSP), but operations on decrypted masked data at the CSP compromise security and increase communication load. To address the challenges of existing cryptography-based privacy-preserving recommendation systems, we propose to exploit Fully Homomorphic Encryption (FHE), which allows arbitrary computations to be performed directly on encrypted data without needing decryption \cite{b27, b28, b29}. FHE maintains privacy throughout the entire recommendation process without additional communication.

A critical challenge in applying FHE to recommendation systems lies in efficiently handling inherently sparse user-item rating matrices. In real-world recommendation scenarios, users typically rate only a small fraction of available items, resulting in matrices where over 90\% of entries are empty \cite{b30}. This sparsity creates significant inefficiencies when naively encrypted, as traditional matrix representations would waste substantial computational resources and bandwidth processing these empty values. This issue will be amplified in FHE, which encrypts data in ciphertexts with much higher space complexity than plaintexts.\par
To address sparsity challenge, we introduce Compressed Sparse Row (CSR) matrix factorization using FHE, which stores only non-zero elements to drastically reduce storage and communication overhead, critical in privacy-preserving systems where each transmitted element incurs substantial costs due to expanded sizes and number of ciphertexts. Moreover, we explore batching techniques from modern FHE schemes like CKKS~\cite{b13}, BFV~\cite{b31}, and BGV~\cite{b32} to reduce computation costs by packing multiple plaintext values into a single ciphertext. Applying batching to sparse matrices is challenging due to their irregular memory access patterns, which conflict with the dense assumptions of conventional methods. Our approach develops a novel technique to integrate batching with CSR representation, enabling efficient parallel processing while maintaining compression and computational efficiency. Our design aligns batching patterns with CSR’s row-oriented structure to maximize ciphertext slot utilization and support FHE optimization.

Our experimental evaluation demonstrates that this integration of CSR with tailored batching techniques achieves substantial improvements in communication efficiency without trading off the accuracy of recommendations while maintaining the privacy guarantees of fully homomorphic encryption, as compared to various state-of-the-art privacy-preserving matrix factorization~\cite{b7},\cite{b11}.\par

This paper is organized as follows: Section ~\ref{sec:II} discusses technical background and related work. Section ~\ref{sec:III} introduces our proposed protocols for FHE-based matrix factorization. Section ~\ref{sec:IV} presents the performance evaluation. Section ~\ref{sec:VI} concludes the paper.


\section{Background}\label{sec:II}

\subsection{Preliminaries}\label{AA}
We propose two protocols: CSR matrix factorization and CSR matrix factorization with optimized batching. Both involve three participants: users, a recommendation server (RS), and a trusted third-party server CSP, as shown in Figure~\ref{fig:FHE_paper1}, where RS performs matrix factorization on encrypted data to predict unrated items; the details are explained in Section~\ref{sec:III}. 
\subsubsection{Recommendation Systems and Collaborative Filtering}

Recommendation systems predict user preferences and provide personalized recommendations. CF assumes that users with similar past preferences will have similar future preferences.\par

In CF, we consider $n$ users and $m$ items with a sparse user-item rating matrix $\mathbb{R}^{n \times m}$. Matrix factorization decomposes this sparse matrix into lower-dimensional user and item latent factor matrices\cite{b15}.
Key notations include: 
\begin{itemize} 
\item $\mathbb{R}$: sparse user-item rating matrix; $[n], [m]$: user and item sets
\item $r_{ij} \in \mathbb{R}$, user $i$'s rating for item $j$
\item ${NZR} \subset [n] \times [m]$: set of non-zero ratings 
\item $M = |{NZR}|$: total non-zero ratings 
\item $k$: latent factor dimension; $\mathbf{U} \in \mathbb{R}^{n \times k}, \mathbf{V} \in \mathbb{R}^{m \times k}$: user/item profiles 
\item $\mathbf{u}_i, \mathbf{v}_j \in \mathbb{R}^k$: latent vectors for user $i$ and item $j$ 
\end{itemize}
Matrix factorization finds profiles $\mathbf{U}, \mathbf{V}$ such that $\mathbb{R} \approx \mathbf{U}\mathbf{V}^T$ with predicted rating $\hat{r}_{ij} = \langle\mathbf{u}_{i}, \mathbf{v}_{j}\rangle$. This solves the regularized optimization, $L(\mathbf{U},\mathbf{V})$:

\begin{equation}
\min_{\mathbf{U},\mathbf{V}}\frac{1}{M}\sum_{(i,j)\in {NZR}}(r_{ij} - \langle\mathbf{u}_i, \mathbf{v}_j\rangle)^2 + \lambda\sum_{i\in[n]}\|\mathbf{u}_i\|_2^2 + \mu\sum_{j\in[m]}\|\mathbf{v}_j\|_2^2
\end{equation}

Where $L(\mathbf{U},\mathbf{V})$ is the loss function we aim to minimize, and $\lambda, \mu > 0$ are regularization parameters to prevent overfitting. Gradient descent iteratively updates profiles: 

\begin{align}
\mathbf{u}_i(t) &= \mathbf{u}_i(t-1) - \alpha\nabla_{\mathbf{u}_i}L(\mathbf{U}(t-1), \mathbf{V}(t-1)), \\
\mathbf{v}_j(t) &= \mathbf{v}_j(t-1) - \alpha\nabla_{\mathbf{v}_j}L(\mathbf{U}(t-1), \mathbf{V}(t-1)),
\end{align}

where $\alpha > 0$ is the learning rate and the gradients are:

\begin{align}
\nabla_{\mathbf{u}_i}L(\mathbf{U}, \mathbf{V}) &= \frac{1}{M}\sum_{j:(i,j)\in{NZR}}\mathbf{v}_j(\langle\mathbf{u}_i, \mathbf{v}_j\rangle - r_{ij}) + \lambda\mathbf{u}_i, \\
\nabla_{\mathbf{v}_j}L(\mathbf{U}, \mathbf{V}) &= \frac{1}{M}\sum_{i:(i,j)\in{NZR}}\mathbf{u}_i(\langle\mathbf{u}_i, \mathbf{v}_j\rangle - r_{ij}) + \mu\mathbf{v}_j.
\end{align}


\subsubsection{Fully Homomorphic Encryption}
Fully homomorphic encryption (FHE) allows arbitrary computations on encrypted data without decryption, yielding the same result as operations on plaintexts \cite{b27}.

Consider two plaintexts $p_1$ and $p_2$ and their corresponding ciphertexts $\llbracket p_1 \rrbracket \leftarrow \text{FHE.Enc}(p_1, pk)$ and $\llbracket p_2 \rrbracket \leftarrow \text{FHE.Enc}(p_2, pk)$. An encryption scheme is additively, and multiplicatively homomorphic if it satisfies:
\begin{align}
p_1 + p_2 = \text{FHE.Dec}(\llbracket p_1 \rrbracket \oplus \llbracket p_2 \rrbracket, sk)\\
p_1 \times p_2 = \text{FHE.Dec}(\llbracket p_1 \rrbracket \otimes \llbracket p_2 \rrbracket, sk),
\end{align}
where $\oplus$ and $\otimes$ represent homomorphic addition and multiplication operators, respectively. FHE schemes also support operations between ciphertexts and plaintexts:
\begin{align}
p_1 + p_2 = \text{FHE.Dec}(\llbracket p_1 \rrbracket \oplus p_2, sk) \\
p_1 \times p_2 = \text{FHE.Dec}(\llbracket p_1 \rrbracket \odot p_2, sk).
\end{align}

\textbf{FHE Schemes:} Several FHE schemes have been developed, each with different characteristics. BGV \cite{b31} and BFV \cite{b32} support exact integer arithmetic, while CKKS \cite{b13} enables approximate real number arithmetic suitable for machine learning. TFHE \cite{b33} provides fast bootstrapping for binary operations. In this work, we use the CKKS scheme for real number approximation, which is well-suited for matrix factorization computations involving floating-point operations.

\textbf{Ciphertext Packing:} Modern FHE schemes support batching, packing multiple plaintext values into a single ciphertext. This technique significantly improves computational efficiency by enabling Single Instruction Multiple Data (SIMD) operations \cite{b24}. CKKS can pack thousands of complex numbers, enabling vectorized operations beneficial for our matrix factorization protocols.

\subsubsection{Compressed Sparse Row Format}

Sparse matrices contain mostly zero entries, making full storage inefficient. The Compressed Sparse Row (CSR) format stores only non-zero elements, reducing memory and computation. As user-item rating matrices are typically sparse, CSR is ideal for recommendation systems.
CSR represents a sparse matrix using three arrays: 
\begin{itemize} 
\item \textbf{Data array} ($\mathbf{data}$): Non-zero values.
\item \textbf{Column indices array} ($\mathbf{col\_indices}$): Column index of each non-zero value. 
\item \textbf{Row pointers array} ($\mathbf{row\_ptr}$): Adjacent pairs indicate row boundaries in the data array, with their difference showing non-zero elements per row. \end{itemize}

\begin{figure}[ht]
\centering
\includegraphics[width=0.5\textwidth]{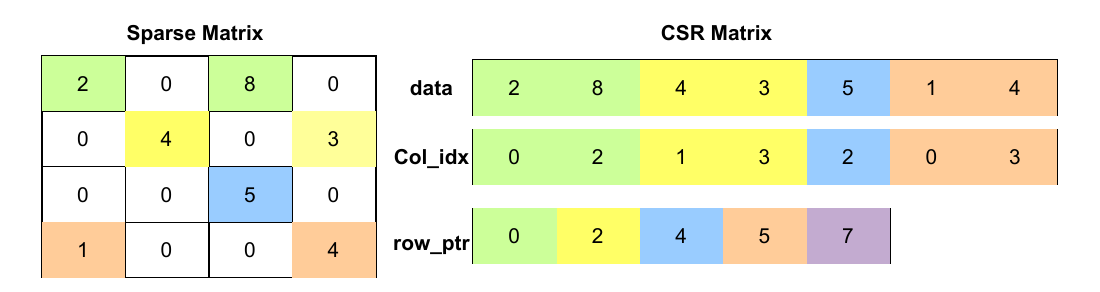}
\caption{Compressed Sparse Matrix Structure.}
\label{fig:saprse_csr}
\end{figure}

For sparse matrix $\mathbb{R}$ with $n$ users, $m$ items, and $M$ non-zero elements, CSR requires only $M$ values, $M$ column indices, and $n+1$ row pointers, compared to $n \times m$ dense storage.
CSR enables efficient row-wise operations fundamental to matrix factorization. Figure~\ref{fig:saprse_csr} shows accessing non-zero value of $i$-th row requires only $\mathbf{data}[\mathbf{row\_ptr}[i] : \mathbf{row\_ptr}[i+1]]$ and corresponding indices. In our FHE-based matrix factorization, CSR reduces required ciphertexts, decreasing communication overhead and computational complexity while maintaining security.

\subsection{Related Work}
According to Rosinosky et al. \cite{b21}, three main approaches achieve privacy-preserving recommendation systems. The decentralized or federated methods \cite{b22}, enabling device collaboration to identify preference similarities without complete data sharing. The obfuscation technique adds strategic noise to mask ratings while maintaining utility, and the homomorphic encryption systems allow computations on encrypted data to preserve user privacy throughout the entire process.\par
\subsubsection{Differential Privacy Based RS}
Differential privacy serves primarily as a non-cryptographic approach to privacy protection \cite{b25}, including in recommendation systems \cite{b26}. This method works by introducing synthetic user ratings to thwart attackers' attempts to infer private user information, while maintaining the overall distribution of system outputs. However, this approach has two significant limitations: it does not secure user data confidentiality from the recommendation system itself, and the accuracy of recommendations tends to decrease as more fake ratings are added to the system.
\subsubsection{Cryptography Based Privacy-Preserving RS}
In 2009, Gentry presented the initial Fully Homomorphic Encryption (FHE) scheme built on ideal lattice structures, enabling unlimited addition and multiplication operations to be performed on encrypted binary data\cite{b16}. Jumonji et al.\cite{b10} developed a privacy-preserving collaborative filtering (PPCF) recommendation system that uses BGV FHE, Smart-Vercauteren (SV) packing, and similarity between users is calculated using cosine similarity. They suggest to offload some homomorphic operations to users and avoiding costly ciphertext–ciphertext multiplications at the server and assumes that participants do not collude. However, user-based CF is less accurate and scalable for large systems with sparse rating matrices, and most real-world recommendation systems rely on latent factor models, for instance, matrix factorization. However, latent factor models usually employ hybrid approach where computationally intensive parts are done on trusted hardware, and only sensitive parts use homomorphic encryption.

Matrix factorization for privacy-preserving recommendation systems was first introduced by Nikolaenko et al. \cite{b7}. The authors used a hybrid approach combining additive homomorphic encryption\cite{b17} with garbled circuits\cite{b18} for secure two-party computation. To handle the sparsity of the rating matrix, they incorporated an oblivious sorting network while maintaining item profiles in unencrypted form. Despite efforts to enhance efficiency through multi-threaded FastGC\cite{b19} and parallel sorting networks, their protocols ultimately resulted in substantial computational and communication overhead. Nayak et al.\cite{b20} improved Nikolaenko et al.'s work by integrating parallelism into the implementation of the oblivious version of graph-based algorithms. Kim et al. \cite{b11} also experimented on further improvement of Nikolaenko et al.'s matrix factorization based model using FHE. Similar to \cite{b7}, in this article, the authors also proposed a three-party protocol between the user, RS, and CSP, and utilized SIMD-style packing\cite{b24} of multiple values into a single ciphertext for batch operations. But some operations, such as division or precision management (e.g., re-scaling and fixed-point normalization), are offloaded to the CSP, holding the secret key. However, the decryption of user data conflicts with the FHE principle.

To the best of our knowledge, our work represents the first privacy-preserving matrix factorization approach combining CSR representation with FHE. Unlike previous methods, our novel framework offers true end-to-end protection by conducting all computational operations entirely in RS within the encrypted domain.

\section{Sparsity-aware FHE Recommendation System}\label{sec:III}

In this work, we propose a novel FHE-based recommendation system to minimize communication overhead and maximize FHE computation throughput, without trading off accuracy and data privacy during recommendation. Specifically, we propose an FHE-only algorithm that exploits data sparsity in user and item matrices during collaborative filtering process.

\subsection{Overview of our Contributions}
FHE enforces strict constraints on ciphertext data structure and operation patterns, posing challenges for recommendation systems. Therefore, we choose the CKKS scheme as it is well-suited as it natively supports real-valued arithmetic operations, which is essential for processing user ratings (e.g., 3.7, 4.5, etc.), and performing gradient descent computations in matrix factorization. CKKS performs well on regular vectors or matrices by exploiting their single-instruction-multiple-data (SIMD) capability, making it non-trivial to transform irregular sparse data structures and operations (e.g., CSR) into efficient FHE implementations. Furthermore, CKKS scheme introduces errors during computation, necessitating an algorithm design that produces accurate recommendation results.

\begin{figure}[htbp]
    \centering
    \includegraphics[scale=0.32]{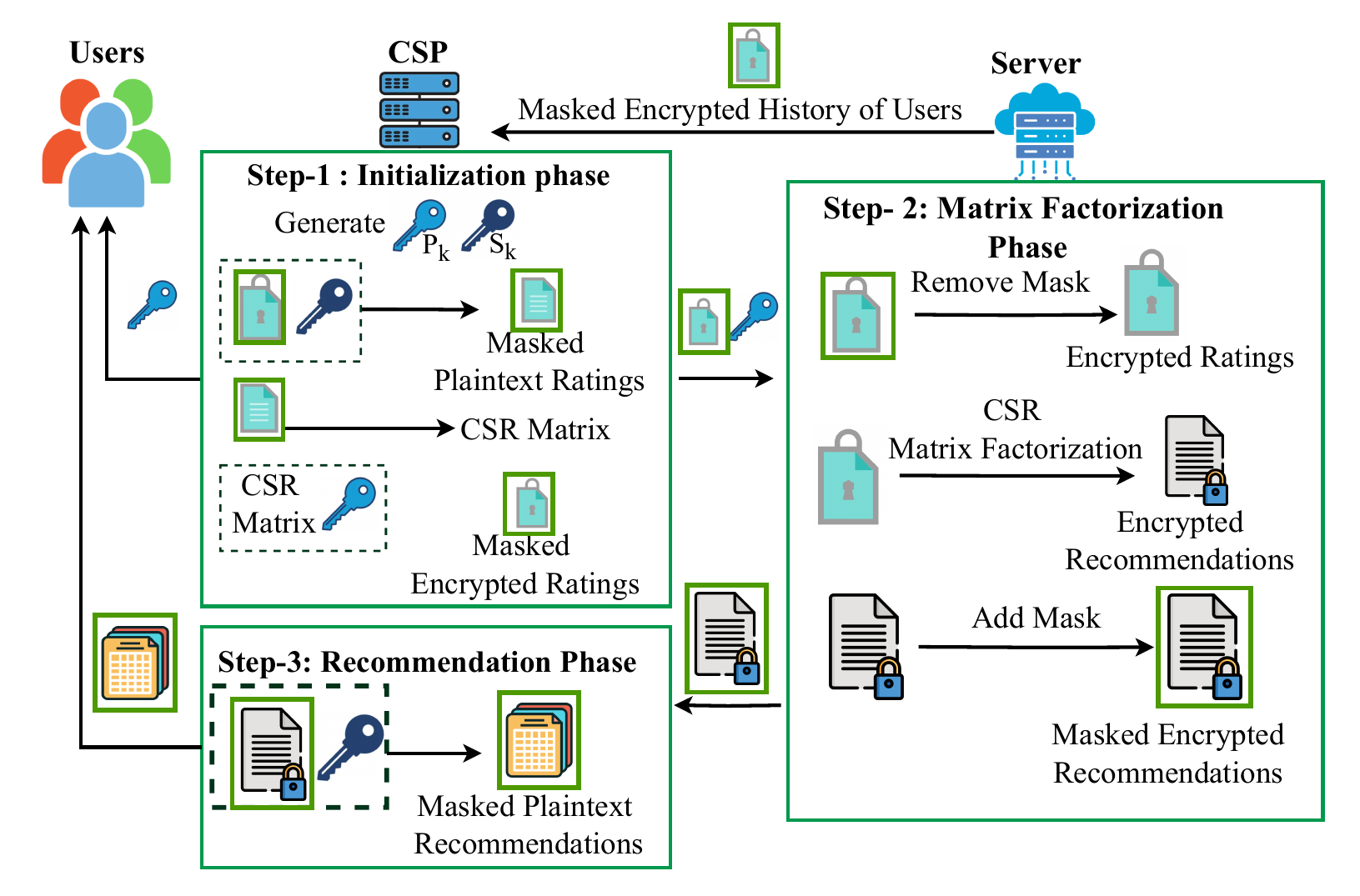}
    \caption{Overview of FHE based CSR matrix factorization.}
    \label{fig:FHE_paper1}
\end{figure}
Our approach introduces two key innovations tackling the challenge mentioned above:\par
\textbf{CSR-FHE Integration:} By adapting Compressed Sparse Row (CSR) matrix representation for FHE-based recommendation systems, we eliminate the need to encrypt and process zero ratings, which constitute the majority of entries in typical recommendation datasets. Compared to prior privacy-preserving recommendation systems, our FHE-only approach addresses communication challenges by minimizing the number of ciphertexts that need transmission and processing. Furthermore, the proposed CSR-FHE algorithm efficiently exploits FHE capability to encrypt various data structures required for sparse computation without harming the accuracy and privacy of collaborative filtering.

\textbf{Optimized Batching Strategy:} We develop an advanced batching optimization technique that enhances computational efficiency by processing multiple user-item profiles as a user-item batch pair while maintaining the benefits of CSR representation. This approach addresses the computational cost challenge by utilizing Single Instruction Multiple Data (SIMD) operations and implementing gradient accumulation strategies, minimizing expensive encryption operations.

\subsection{FHE-only Privacy-Preserving Recommendation System}
Figure~\ref{fig:FHE_paper1} shows the system for our proposed CSR matrix factorization approach with FHE, consisting of three phases: initialization, matrix factorization, and recommendation. 

\subsubsection{Initialization Phase}
In the initialization phase, CSP generates a key pair ($pk$, $sk$), and distributes public key $pk$ to users and RS. Users encrypt their rating with $pk$ and send it to RS. RS adds mask to encrypted ratings and shares masked encrypted rating with CSP. CSP then
\begin{itemize}
\item Decrypts masked encrypted ratings with private keys $sk$ and get masked plaintext ratings
\item Converts masked plaintext rating into a masked CSR matrix  
\item Encrypts masked CSR matrix with $pk$, and sends to RS
\end{itemize}
 
\subsubsection{Matrix Factorization Phase}
This phase is performed entirely in RS, where user profile matrix $U$ and item profile matrix $V$ are randomly generated and encrypted with public key $pk$. RS removes the mask from the masked encrypted CSR matrix and performs CSR matrix factorization in a fully encrypted domain using gradient updates. Each encrypted rating updates encrypted user-item profiles by minimizing error and generates an encrypted predicted ratings matrix. Finally, the RS add mask with the predicted rating matrix, and sends the masked encrypted recommendation to the CSP.

\subsubsection{Recommendation Phase}
CSP decrypts the masked predicted rating matrix with private key and sends plaintext masked recommendations to users. Since mask values remain exclusively known to RS and users, neither RS nor CSP ever gains access to actual plaintext ratings throughout the process. This dual-layer protection—encryption combined with masking—ensures the security of the process.

\subsection{CSR Matrix Factorization using FHE}
Matrix factorization is the main computational task in the proposed secure recommendation system.
A naive way to perform gradient descent-based matrix factorization using FHE would process $n\cdot m$ ciphertexts for matrix factorization for sparse matrix $\mathbb{R}$, where $n$ and $m$ are the number of users and items, respectively. 
Even with SIMD packing and encryption of each user ${u}_i$'s ratings, this generates $n$ ciphertexts for ${n}$ users. 
If the communication cost between user, RS, and CSP is defined by the number of ciphertext exchanges as $\mathbb{CT}$, naive approach would generate (${n+1}$) $\mathbb{CT}$  for sending ${n}$ ciphertexts from the CSP to the RS and 1 ciphertext for the encrypted predicted rating matrix from the RS to the user.

To overcome inefficiencies of naive approaches that process all matrix elements including zeros, we propose an elaborate data structure that fully exploits available ciphertext slots using CSR matrix representation. This optimization reduces protocol's communication cost from $(1+n)$ to only $(1+M/L)$ $\mathbb{CT}$, where $L$ is slots per ciphertext, and $M$ is the number of non-zero ratings generated by $n$ users, providing significant improvements in efficiency without compromising privacy.
\begin{figure}[ht]
\centering
\includegraphics[scale=0.45]{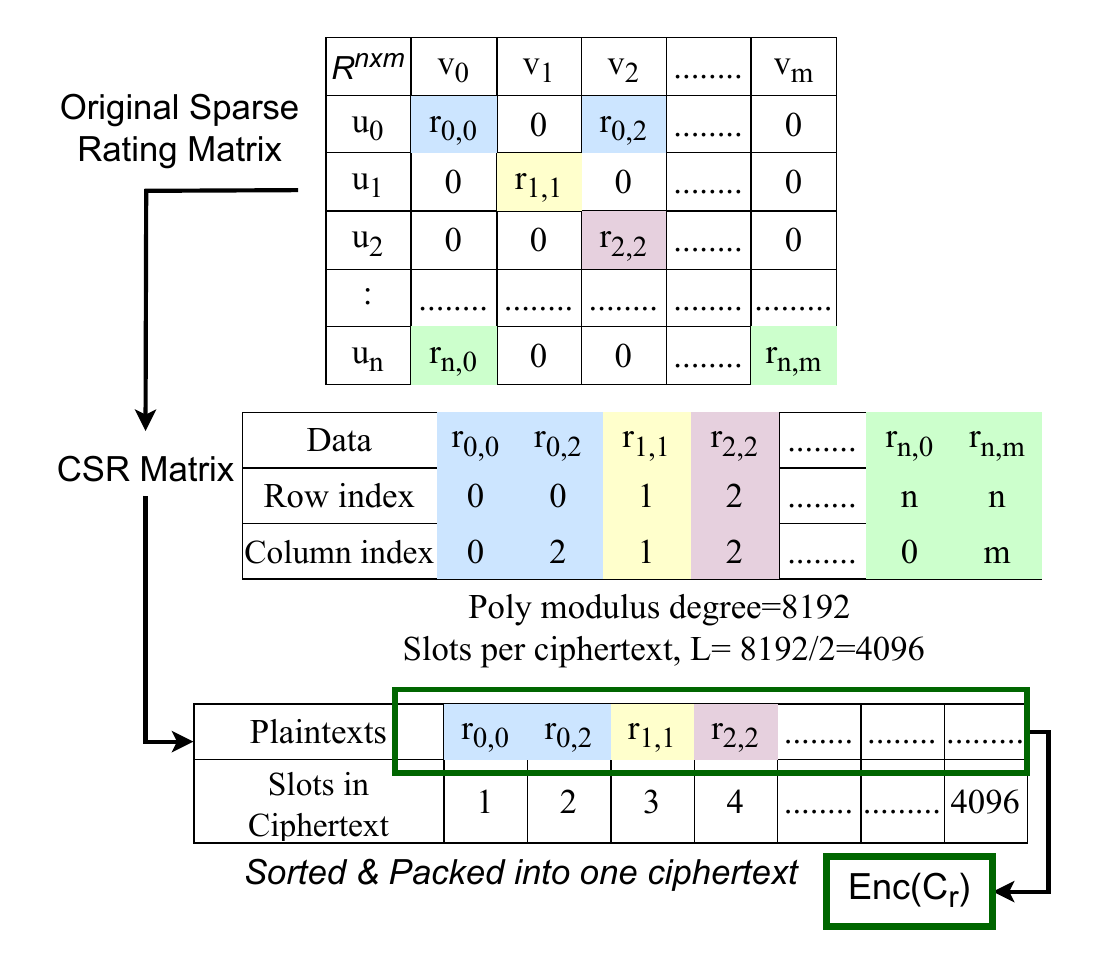}
\caption{Packing of CSR matrix.}
\label{fig:CSR_FHE}
\end{figure}

\begin{algorithm}[htb!]
\caption{CSR Matrix Factorization using FHE}
\label{alg:encrypted_mf}
\begin{algorithmic}[1]
\Require Sparse rating matrix, $\mathbb{R}^{m \times n}$, number of factors $k$, learning rate $\alpha$, regularization parameter $\lambda$, number of iterations $T$, encryption context 
\Ensure Encrypted predicted rating matrix $\hat{R}$

\State \textbf{\#EncryptCsrMatrix (executed on CSP):}
\State $(row\_indices, col\_indices) \gets \text{nonzero}(R)$
\State $values \gets R.data$
\State $encrypted\_data \gets \text{CKKS.Encrypt}(values)$

\State \textbf{\#CsMatrixFactorization (executed on RS):}
\State Initialize $U \in \mathbb{R}^{m \times k}$ with random values $\sim \mathcal{U}(0,1)$
\State Initialize $V \in \mathbb{R}^{n \times k}$ with random values $\sim \mathcal{U}(0,1)$

\State $Enc\_U \gets [\text{Encrypt}(U[i]) \text{ for } i \in \{0,1,\ldots,m-1\}]$
\State $Enc\_V \gets [\text{Encrypt}(V[j]) \text{ for } j \in \{0,1,\ldots,n-1\}]$

\For{$t = 1$ to $T$}
    \For{\text{each $Enc\_data$ in CSR matrix}}
        \State $user \gets row\_indices[idx]$
        \State $item \gets col\_indices[idx]$
        
        \State $error \gets Enc\_data[idx]-(Enc\_U_{user} \cdot Enc\_V_{item}^{\top})$
        
        \State $GD\_U \gets (error \cdot Enc\_V_{item}) - (\lambda \cdot Enc\_U_{user})$
        \State $GD\_V \gets (error \cdot Enc\_U_{user}) - (\lambda \cdot Enc\_V_{item})$
        
        \State $Enc\_U[user] \gets Enc\_U[user] + \alpha \cdot GD\_U$
        \State $Enc\_V[item] \gets Enc\_V[item] + \alpha \cdot GD\_V$
    \EndFor
\EndFor
    
\State $\hat{R} \gets Enc\_U \cdot Enc\_V^{\top}$\\
\Return $\hat{R}$
\end{algorithmic}
\end{algorithm}

Our CSR-FHE approach is detailed in Algorithm~\ref{alg:encrypted_mf}, which consists of CSR matrix encryption and encrypted matrix factorization. In $EncryptCsrMatrix$ function, non-zero elements are extracted from sparse rating matrix $\mathbb{R}$, and only actual rating values are encrypted in $encrypted\_data$ in CSR format, significantly reducing the data volume. Because of CSR extraction, $encrypted\_data$ contains sorted ratings by user ids.  Figure~\ref{fig:CSR_FHE} illustrates the extraction of CSR from sparse matrix $\mathbb{R}$, and packing of sorted non-zero ratings. With polynomial modulus degree 8192, we achieve $L = 4096$ plaintext slots, allowing 4096 user-item ratings per ciphertext. This reduces communication by factor $L$, requiring only $M/L$ ciphertexts. For ratings $r_{i,j} \in{NZR}$, we pack them as:\\
$
C_{r_1} = r_{i_1,j_1} \parallel r_{i_2,j_2} \parallel \ldots \parallel r_{i_L,j_L}
$
where $\parallel$ denotes concatenation, and each $r_{i,j}$ represents a unique user-item pair. The core $CsMatrixFactorization$ function in Algorithm~\ref{alg:encrypted_mf}, RS performs gradient descent entirely in the encrypted domain. For the number of latent factors $k$, randomly generated profile matrices $U$ and $V$ have $k$ elements in each row. 
\begin{figure}[ht]
\centering
\includegraphics[scale=0.39]{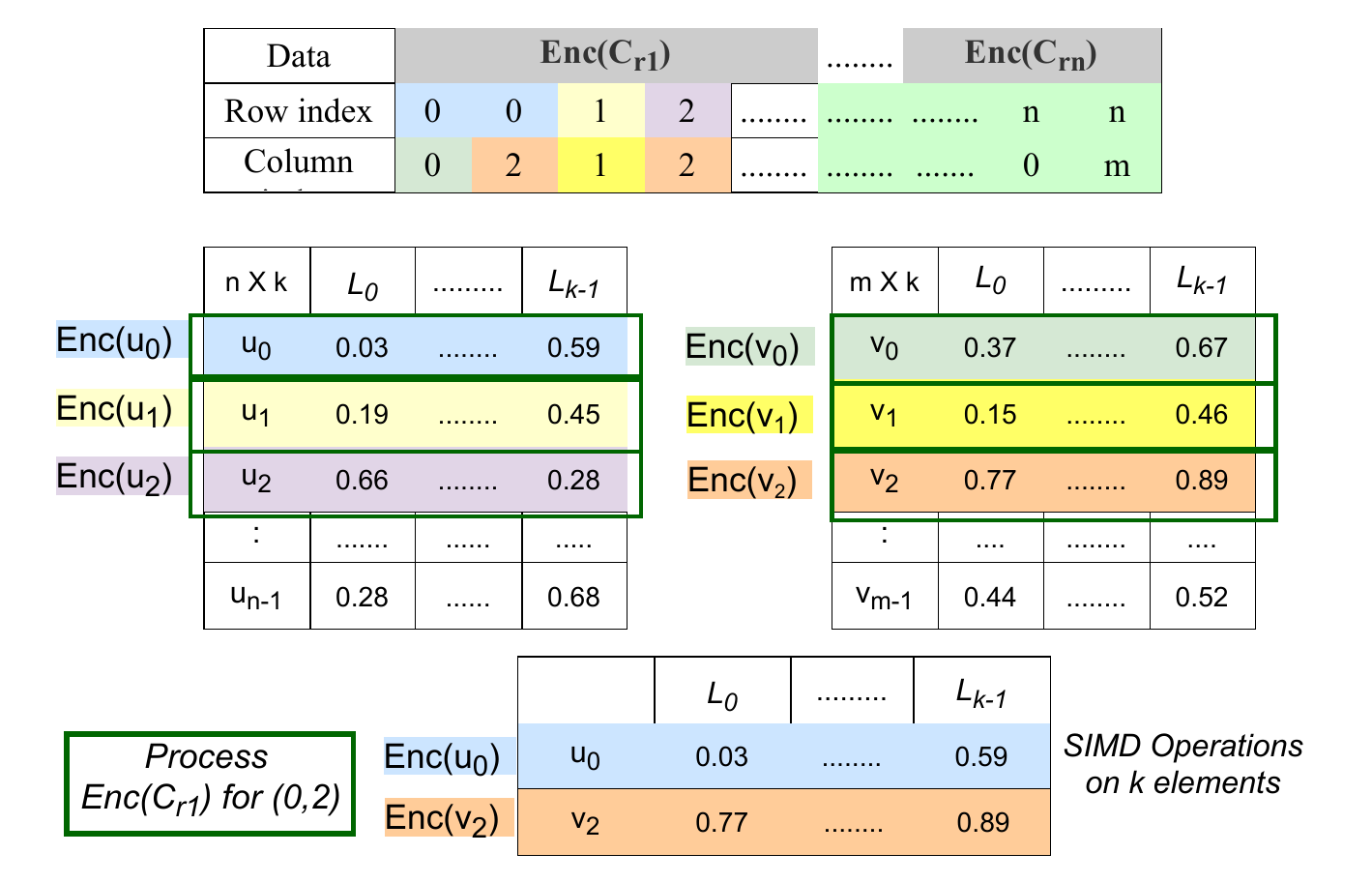}
\caption{FHE based CSR matrix factorization.}
\label{fig:CSR_Matrix_Factorization}
\end{figure}

As shown in Figure~\ref{fig:CSR_Matrix_Factorization}, all the independent elements in a profile matrix, thus $k$ elements of user/item profile vectors, are packed into single ciphertexts.

Let $\text{Enc}(ui)$ and $\text{Enc}(vj)$ for $u_i = (u_{L0} \parallel \ldots \parallel u_{Lk-1})$ and $v_j = (v_{L0} \parallel \ldots \parallel v_{Lk-1})$; where $L_{0}$ to $L_{k-1}$ denotes $k$ dimensional latent factors.\par
The $Enc\_U$ and $Enc\_V$ packed ciphertext of Algorithm~\ref{alg:encrypted_mf} are then processed for all the values in $Enc\_Data$ of CSR matrix, where the $GD_U$ and $GD_V$ store gradients to process only non-zero ratings and their corresponding user-item profile updates. Finally, encrypted predicted rating matrix $\hat{R}$ is calculated by the updated encrypted user and item profile matrices.\par

\subsection{CSR with Optimized Batching}
While our CSR-FHE approach addresses communication challenges, further optimization is needed for computational efficiency. Although it packs all independent elements of each user/item matrix into a single ciphertext, it does not fully utilize the benefits of SIMD packing when processing only one user and item profile at a time.

To address the limitation, we propose an integrated approach that combines CSR representation with efficient batching strategies. In this architecture, for the non-zero ratings in the CSR matrix, only the corresponding user and item profiles are updated, eliminating unnecessary computations.
\begin{figure}[htbp]
    \centering
    \includegraphics[scale=0.4]{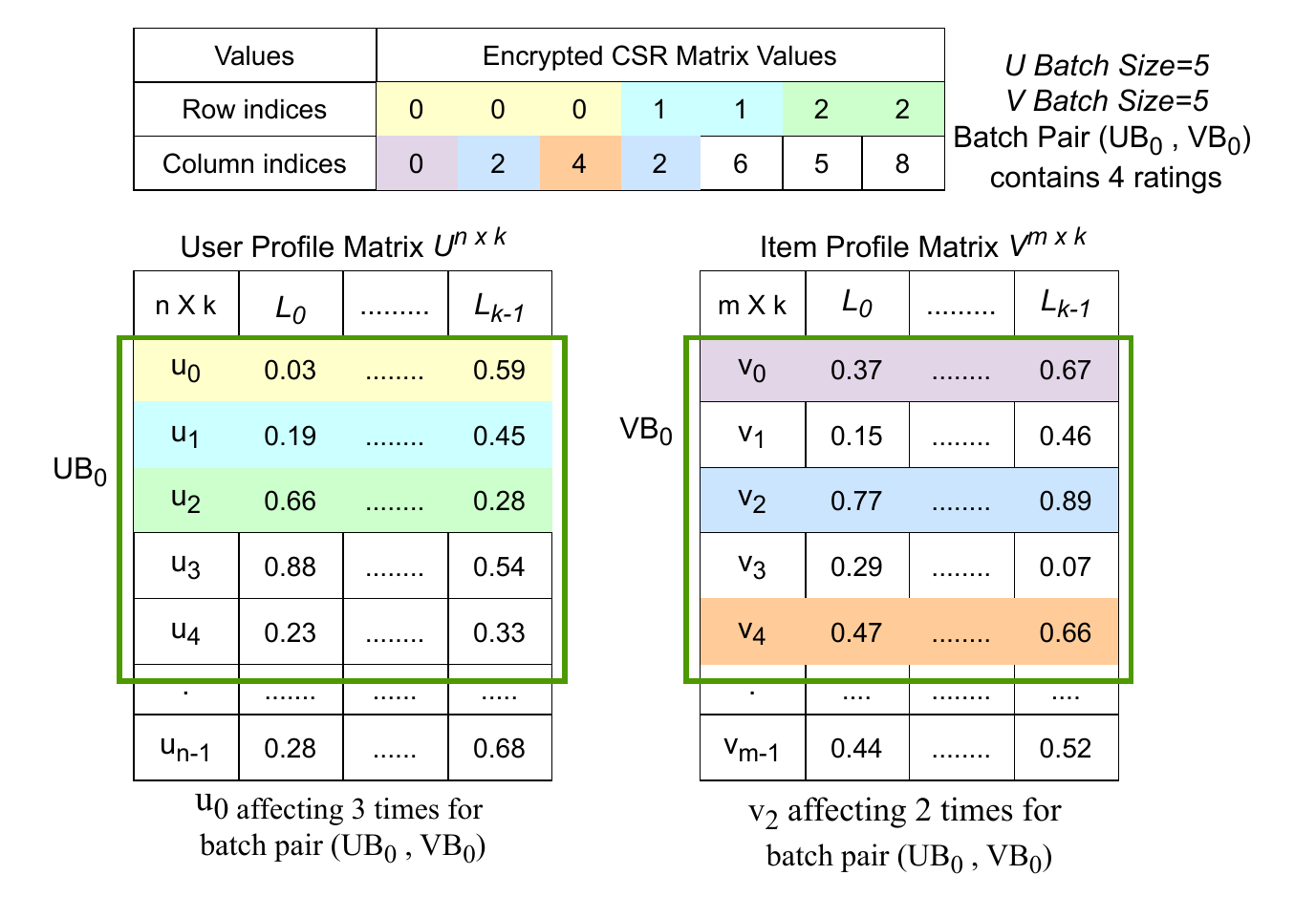}
    \caption{User-item batch pair generation.}
    \label{fig:BatchPairGeneration}
\end{figure}

\begin{algorithm}[htb!]
\caption{Gradient Update with Accumulator}
\label{alg:gradient_update}
\begin{algorithmic}[1]
\Require Batch groups $batch\_groups$, encrypted factors $U\_batches$, $V\_batches$
\Ensure Updated factor batches

\For{\textbf{each} $(u\_batch\_idx, v\_batch\_idx), ratings \in batch\_groups$}
    \State $Enc\_U\_batch \gets Enc\_U\_batches[u\_batch\_idx]$ 
    \State $Enc\_V\_batch \gets Enc\_V\_batches[v\_batch\_idx]$
   
    \State $Enc\_U\_updates \gets \{\}$ 
    \State $Enc\_V\_updates \gets \{\}$ \text{Initialize accumulators}
    
    \For{\textbf{each} $(i, user, item) \in ratings$}
       
        \State $Enc\_u\_factors \gets Enc\_U\_batch[u\_local\_idx]$ 
        \State $Enc\_v\_factors \gets Enc\_V\_batch[v\_local\_idx]$
        
        \State $error \gets rating[i] - (u\_factors \cdot v\_factors^{\top})$
        \text{Dampened learning rate}
        \State $\alpha' \gets \alpha / (1 + 0.1 \cdot iteration)$ 
        
        \State $grad\_u \gets \alpha' \cdot (error \cdot Enc\_v\_factors - \lambda \cdot Enc\_u\_factors)$
        \State $grad\_v \gets \alpha' \cdot (error \cdot Enc\_u\_factors - \lambda \cdot Enc\_v\_factors)$
        
        \State $Enc\_U\_updates[u\_local] \gets Enc\_U\_updates.get(u\_local) + grad\_u$ \text{Accumulate}
        \State $Enc\_V\_updates[v\_local] \gets Enc\_V\_updates.get(v\_local) + grad\_v$
    \EndFor
    
    \State Apply all accumulated updates in $Enc\_U\_updates$ to $Enc\_U\_batches[u\_batch\_idx]$
    \State Apply all accumulated updates in $Enc\_V\_updates$ to $Enc\_V\_batches[v\_batch\_idx]$
\EndFor

\Return $Enc\_U\_batches, Enc\_V\_batches$
\end{algorithmic}
\end{algorithm}

Figure~\ref{fig:BatchPairGeneration} illustrates our user-item batch pair generation strategy. For optimized batching, we group ratings by their corresponding user and item batch indices, ensuring only profiles associated with actual ratings are processed. For encrypted CSR matrix values in Figure~\ref{fig:BatchPairGeneration}, let us consider a subset of encrypted CSR matrix values with row-column index set $S = \{(0, 0), (0, 2), (0, 4), (1, 2), (1, 6), (2, 5), (2, 8)\}$ and number of users, items, and latent factor is $n, m,$ and $k$ respectively. User/item profile matrics are initially generated with random values (0 to 1). Row indices (local user indices) are inherently sorted due to CSR matrix structure. Assuming user/item batch sizes of 5, we define the following batches:\\
User profile batches,\\
$\mathbf{UB}_0 = (u_0 \parallel u_1 \parallel u_2 \parallel u_3 \parallel u_4)$: user batch index 0,
$\mathbf{UB}_1 = (u_5 \parallel u_6 \parallel u_7 \parallel u_8 \parallel u_9)$: user batch index 1,\\
Similarly, item profile batches,\\
$\mathbf{VB}_0 = (v_0 \parallel v_1 \parallel v_2 \parallel v_3 \parallel v_4)$: $v_0$ local index of item 0
$\mathbf{VB}_1 = (v_5 \parallel v_6 \parallel v_7 \parallel v_8 \parallel v_9)$

Our approach groups all ratings in set $S$ by user-item batch pairs. For instance, the batch pair $(\mathbf{UB}_0, \mathbf{VB}_0)$ contains 4 ratings, while $(\mathbf{UB}_0, \mathbf{VB}_1)$ contains 3 ratings. We then process each rating group collectively, performing a single gradient update after processing all ratings within a group.

Another key innovation of our CSR optimized batching lies in our gradient accumulation strategy, implemented in Algorithm~\ref{alg:gradient_update}. Rather than updating gradients individually for each rating, we accumulate all gradient updates for each user/item within a batch pair and apply them collectively. $Enc\_U\_updates$ and $Enc\_V\_updates$ are batch update accumulators holding user/item local index as key and corresponding $grad\_u$/$grad\_v$ as values. For instance, in Figure~\ref{fig:BatchPairGeneration}, the specific case of batch pair $(\mathbf{UB}_0, \mathbf{VB}_0)$, user local index 0 appears three times. Rather than updating gradient for $u_0$ three separate times, we store all gradients for each local index in a dictionary structure—a gradient accumulator—and update each local index once per batch pair. This strategy significantly reduces the number of encryption operations required, thereby enhancing computational efficiency while maintaining the privacy guarantees of homomorphic encryption.

\section{Experimental Result Analysis}\label{sec:IV}

\subsection{Experimental Setup}
We evaluated our FHE-based CSR matrix factorization and CSR with optimized batching methods using Python, and TenSEAL library~\cite{b12}, built on Microsoft SEAL~\cite{b35}. Our privacy-preserving matrix factorization framework was implemented and evaluated using Google Colaboratory's CPU environment with Intel Xeon processors at 2.3 GHz with 12.7 GB RAM. We employed the CKKS scheme~\cite{b13} for leveled homomorphic encryption over real numbers. Experiments used MovieLens 100K dataset~\cite{b14} (100,000 ratings, 943 users, 1,682 movies). For comparison with prior work of Nikolaenko et al. \cite{b7}, and Kim et al. \cite{b11}, we evaluate computational time and communication performance for the dataset restricted to 40 most popular movies. The restricted dataset contains 14K ratings generated by 940 users. Experiments use subsets of \{128, 256, 512, 1024\} non-zero ratings from the dataset\cite{b14} to align with the same rating sets of Nikoleanko et al.\cite{b7}, and Kim et al.\cite{b11} for experimental analysis. We set profile dimension $k=10$, fractional precision 32 bits, and polynomial modulus degree $2^{13}=8192$ with $128$ bit security. 

\subsection{Evaluation Metrics}
We evaluated our methods using the following metrics:
\par\textbf{Computational Efficiency:}
We measured execution time for gradient descent updates of encrypted user and item profiles, excluding encryption/decryption operations which are significantly faster.

\textbf{Communication Overhead:}
We quantified total bytes transmitted between target user and server during initialization, matrix factorization, and recommendation phases.

\textbf{Accuracy:}
We measured the prediction accuracy using Root Mean Squared Error (RMSE):

\begin{equation}
\text{RMSE} = \sqrt{\frac{\sum_{(u,v) \in {NZR}}(\hat{r}_{uv} - r_{uv})^2}{|{NZR}|}}
\end{equation}

\noindent where $\hat{r}_{uv}$ represents the rating predicted by our encrypted methods and $r_{uv}$ is the corresponding plaintext rating by user $u$'s preference on item $v$ in the non-zero rating set ${NZR}$. Lower RMSE values indicate higher prediction accuracy. Our solution introduces minor approximation errors due to floating-point representation in the CKKS scheme.

\subsection{Comparison to Prior Privacy-Preserving Methods}

\begin{figure}[ht]
\centering
\includegraphics[scale=0.14]{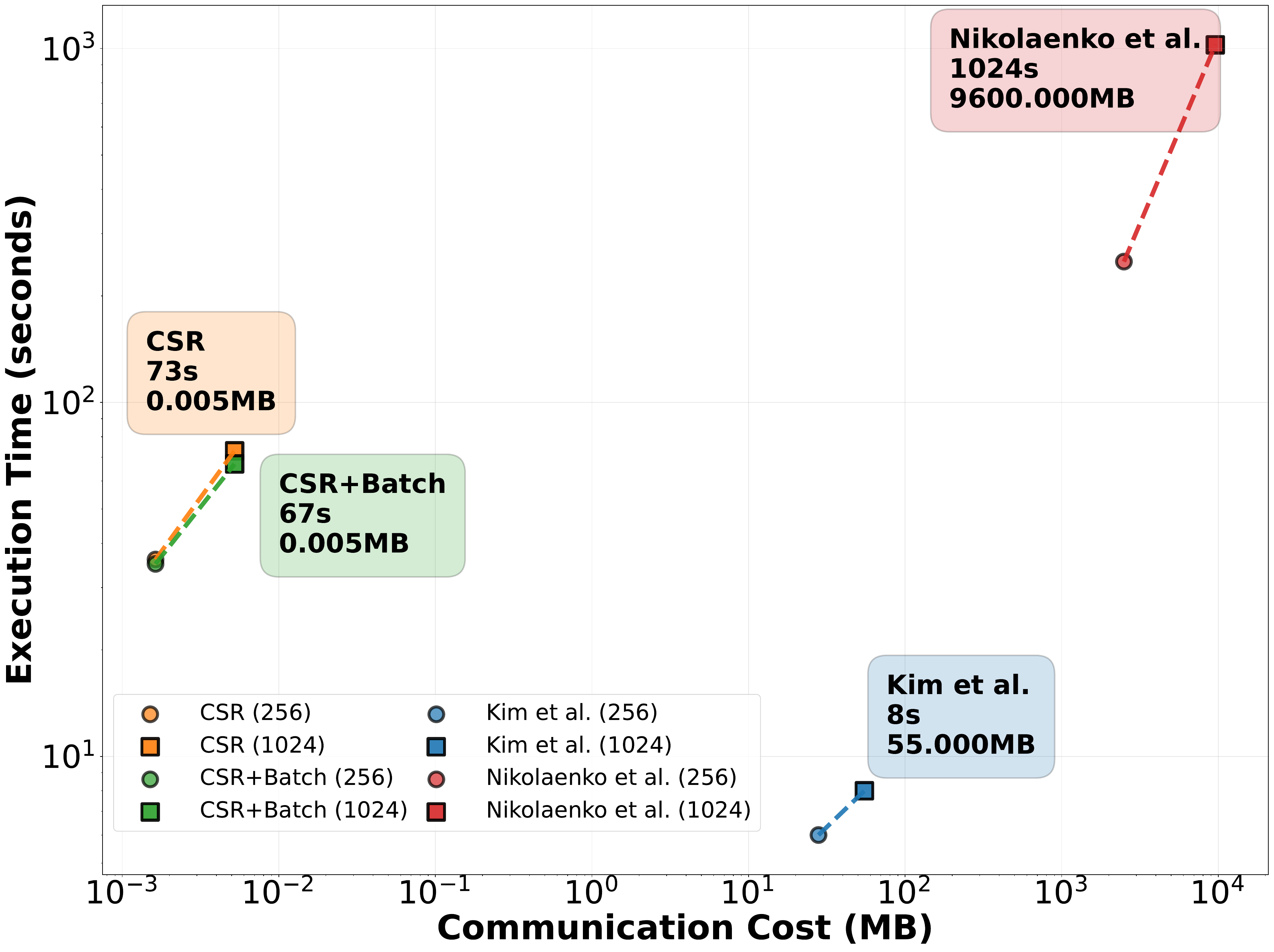}
\caption{Comparison of Performance Trade-off Analysis.}
\label{fig:performance_scatter_plot}
\end{figure}

Figure~\ref{fig:performance_scatter_plot} shows computational and communication overhead performance of our approaches and prior methods: Nikoleanko et al.\cite{b7}, and Kim et al.\cite{b11}, using the same number of non-zero ratings. Dashed lines connect performance from 256 to 1024 ratings; bottom left indicates better performance. \par
The y-axis of Figure~\ref{fig:performance_scatter_plot} illustrates per-iteration execution time for matrix factorization updates. Since each measurement reflects one iteration, total runtime scales linearly with the number of iterations
$T$. In Nikolaenko et al. \cite{b7}, garbled circuits evaluation leads to excessive computation and communication overhead for arithmetic-intensive matrix factorization.
Our FHE-only methods achieve 
significantly lower computation time compared to Nikoleanko et al.\cite{b7} because FHE can directly compute arithmetic operations without expensive logic-level evaluation. Additionally, matrix factorization in FHE is performed entirely within the encrypted domain, eliminating any communication overhead during this phase. Kim et al. \cite{b11} achieve lowest computation time because their approach performs most computations on plaintexts masked by MPC protocol. 
However, even for 256 ratings, their approach requires tens of megabytes of data transferred per iteration, resulting in prohibitively high communication overhead. 

As shown on the x-axis of Figure~\ref{fig:performance_scatter_plot}, both of our proposed methods incur the same and lowest communication cost compared to previous approaches of Nikoleanko et al.~\cite{b7}, and Kim et al.~\cite{b11}. Our methods require communication only once during the entire process (during initialization and recommendation phases), while prior works required repeated communication at every iteration of their matrix factorization phase. For sparse datasets like MovieLens 100k\cite{b14}, typically a minimum of 20 iterations are required to achieve significant accuracy improvements\cite{b36}, meaning prior approaches incur a substantial communication overhead of at least 20 times throughout the process. Our matrix factorization operates entirely within RS with zero communication during this phase, while previous works involve substantial ciphertext exchange proportional to iterations required for convergence. Furthermore, considering recent advancements in hardware acceleration for FHE, which achieves more than 5 orders of magnitude speedup over conventional CPU systems~\cite{ufc,sharp,f1,fhemem}, our FHE-only approach shows significant advantages over prior methods due to much less communication overhead. \par
Notably, Figure~\ref{fig:performance_scatter_plot} highlights that with increasing non-zero ratings, our proposed CSR with optimized batching outperforms CSR without batching. While both methods take almost the same computational time for 256 ratings, when sparsity decreases to 1024 ratings, CSR with optimized batching performs approximately 8\% faster than CSR without batching. We note that both CSR methods are highly parallelizable (across users or items), so execution time can be further improved via parallel implementation. Section~\ref{sec:V} shows detailed analysis of batching optimization for our CSR-FHE algorithms.

\subsection{Accuracy Analysis}
\begin{figure}[ht]
\centering
\includegraphics[scale=0.35]{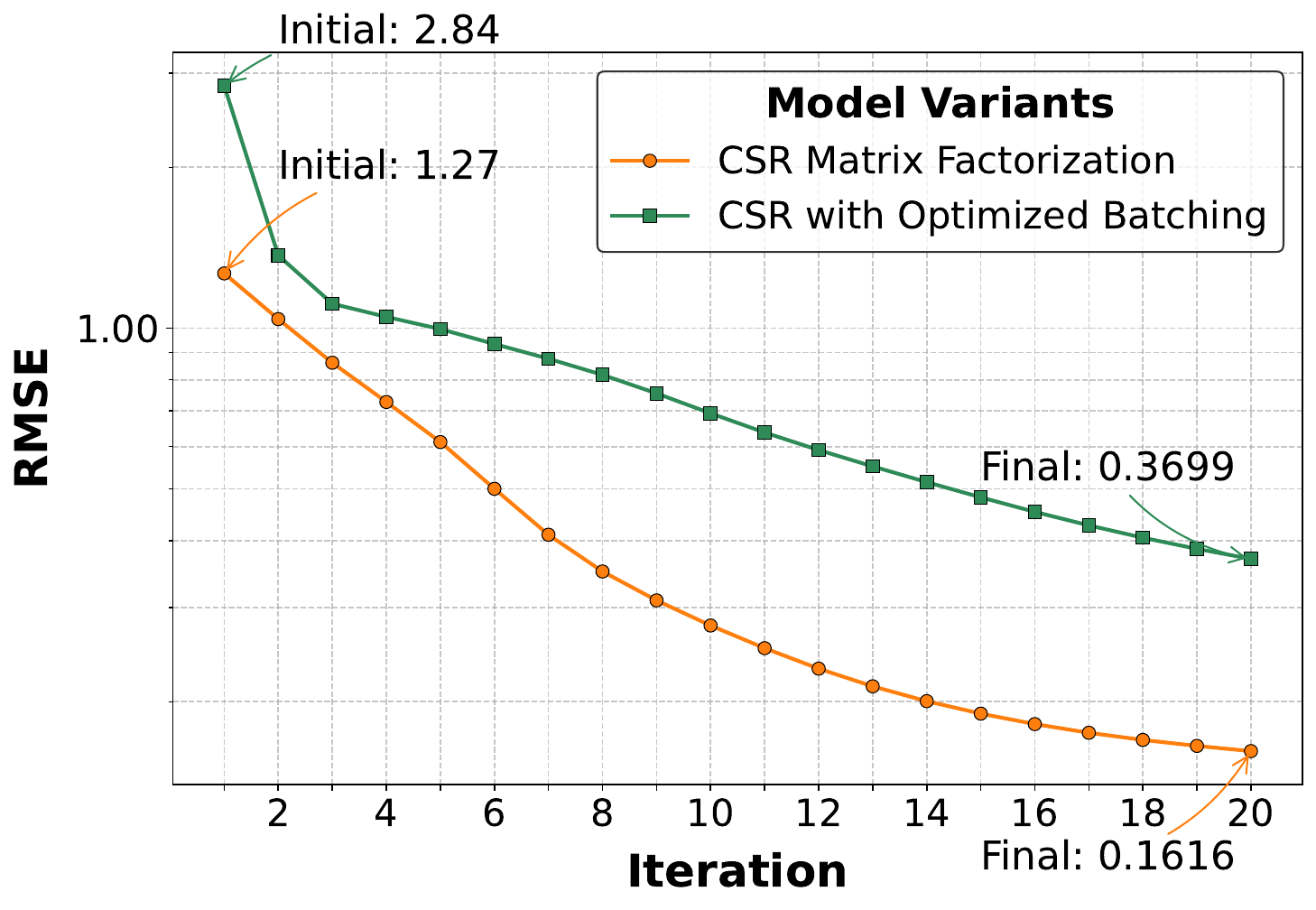}
\caption{RMSE over iterations for CSR Matrix Factorization variants: with and without batching (32-bit fractional precision).}
\label{fig:rmse_csr_variants}
\end{figure}

\begin{figure}[ht]
\centering
\includegraphics[scale=0.35]{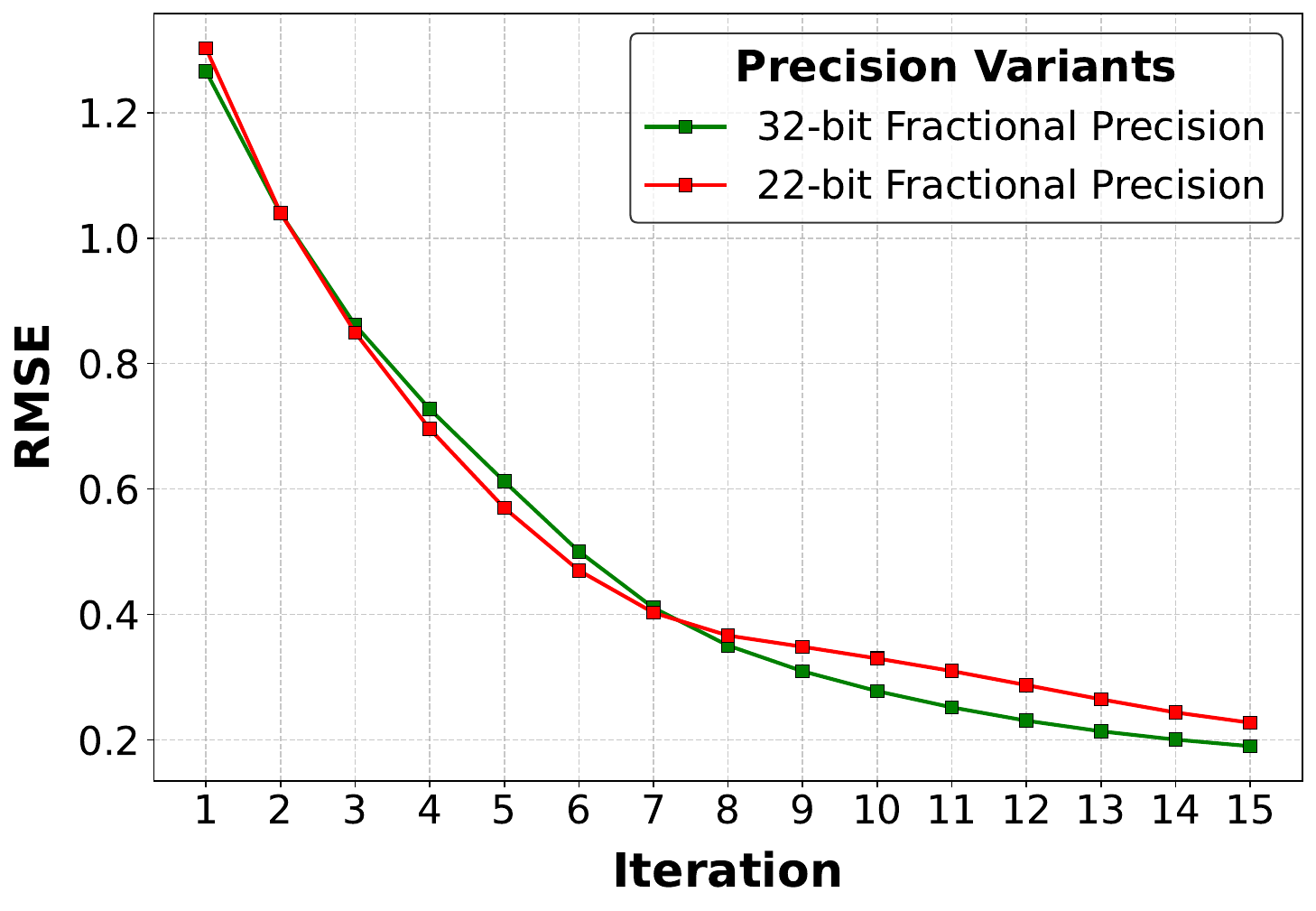}
\caption{Effect of fractional precision (22-bit vs.\ 32-bit) on RMSE over 15 iterations of CSR Matrix Factorization without batching.}
\label{fig:rmse_precision_effect}
\end{figure}

We study the convergence behavior of our CSR-based matrix factorization methods under different FHE parameterizations, particularly focusing on bits allocated to the fractional part of the CKKS scheme and gradient descent iterations. With Poly modulus degree 8192 and bit scale 32, Figure~\ref{fig:rmse_csr_variants} illustrates the RMSE evolution over 20 iterations for CSR without batching and CSR with optimized batching. Both methods achieve approximately 87\% reduction in RMSE. Notably, the CSR method without batching converges approximately 2.29 times faster. 
Furthermore, Figure~\ref{fig:rmse_precision_effect} shows CKKS fractional precision impact on convergence. Using the CSR method without batching, we observe that increasing fractional bit precision from 22 to 32 bits results in consistently lower RMSE across iterations. This demonstrates that higher fractional precision enables better numerical stability and more accurate updates during encrypted gradient descent operations.\par
\subsection{Overall Performance Comparison}

\begin{figure}[ht]
\centering
\includegraphics[scale=0.33]{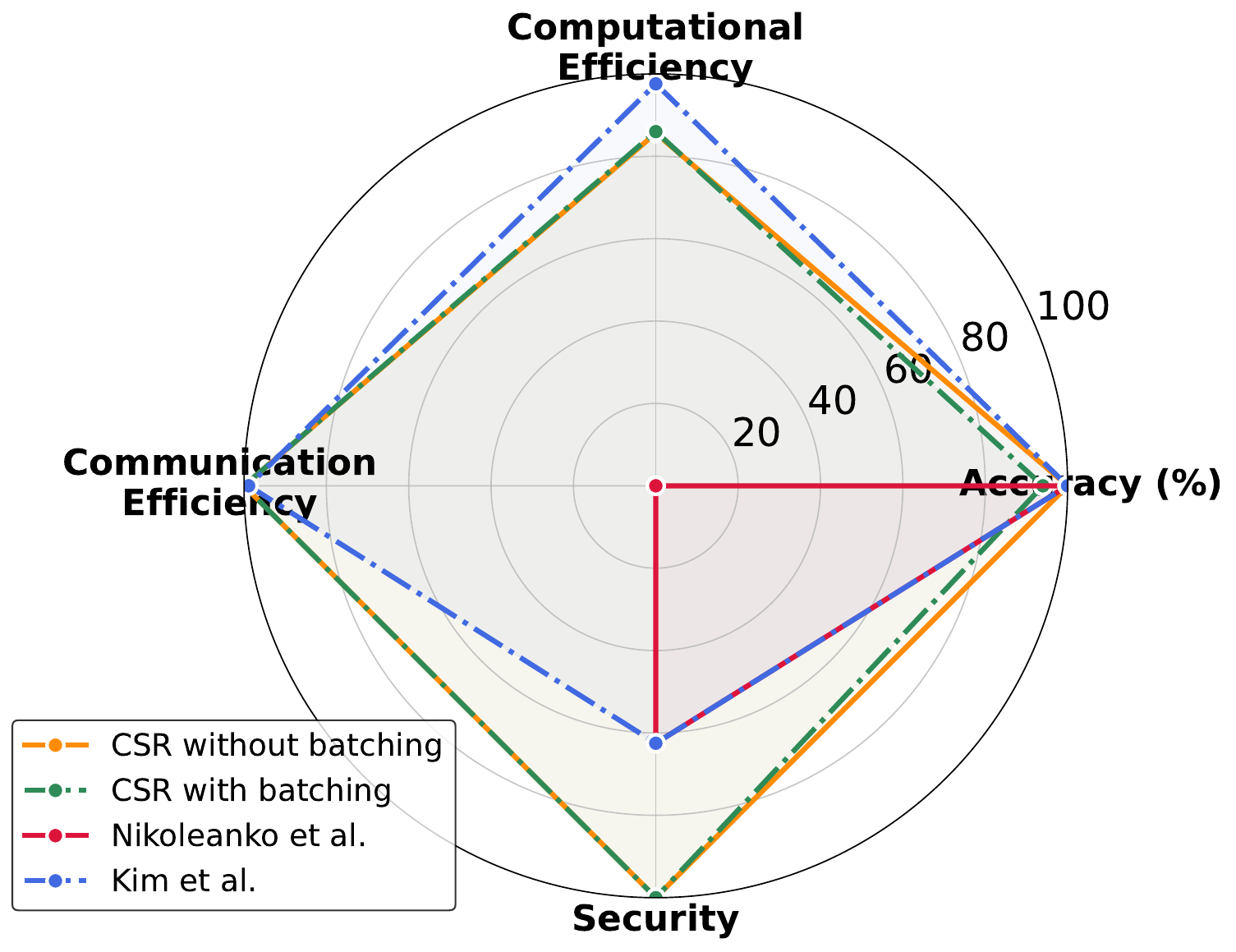}
\caption{Performance metrics comparison (CSR methods is the most balanced).}
\label{fig:Overall_performance_comparison}
\end{figure}
Figure ~\ref{fig:Overall_performance_comparison} illustrates the per-iteration performance evaluation compared to our CSR-based approaches against existing methods from Nikoleanko et al.\cite{b7}, and Kim et al.\cite{b11} for 256 ratings. The CSR methods provide enhanced security with 128-bit encryption compared to 80-bit security level. Kim et al.'s\cite{b11} method demonstrates superior computational performance at 6 seconds but incurs substantial communication overhead of 28MB. Conversely, Nikoleanko et al.'s\cite{b7} approach exhibits slowest performance with 250 seconds computation time for each iteration and 2.5GB communication requirements. Regarding communication efficiency, CSR-based methods significantly outperform existing solutions with the lowest overhead. Overall, CSR with optimized batching approach delivers optimal trade-offs across security, computation, and communication metrics.

\subsection{Analysis between CSR and CSR with Batching}\label{sec:V}

Both proposed approaches have identical communication costs since they exchange the same number of ciphertexts ($\mathbb{CT}$) during initialization and recommendation phases. However, CSR with batching achieves superior computational efficiency by processing multiple ratings simultaneously rather than individually. While the difference is minimal for sparse datasets, the advantage becomes pronounced as ratings increase. The computational complexity is dominated by gradient descent updates on fully encrypted data. Table~\ref{tab:complexity} presents the complexity analysis for our models, where $n$ represents the number of users, $m$ number of items, $k$ dimensionality of user/item profiles, and $M$ the number of non-zero ratings in the system.

\begin{table}[htbp]
\centering
\caption{Complexity of CSR-based Methods}
\label{tab:complexity}
\resizebox{\columnwidth}{!}{
\begin{tabular}{@{}lll@{}}
\toprule
\textbf{Operation} & \textbf{CSR Matrix Factorization} & \textbf{CSR with Optimized Batching} \\ 
\midrule
Initialization & $\mathcal{O}((m  + n) \times k)$ & $\mathcal{O}((m/u\_batch\_size + n/v\_batch\_size) \times k)$ \\
Training Loop & $\mathcal{O}(T \times M \times (k + E))$ & $\mathcal{O}(T \times M \times (k + E_{batch}))$ \\

Update Operations & $\mathcal{O}(T \times M \times E)$  & $\mathcal{O}(T \times M \times E_{batch})$  \\
\bottomrule
\end{tabular}}
\end{table}

In this analysis, $E$ represents homomorphic operation costs for individual user-item profiles, while $E_{batch}$ denotes batch operation costs. When multiple updates affect the same user/item profile within a single batch operation, $E_{batch}$ achieves lower computational cost by reducing redundant encryption operations. However, as sparsity levels increase within batches, the reduction in execution time becomes less pronounced. Nevertheless, the CSR format combined with optimized batching strategy significantly decreases the overall computational burden by enabling simultaneous processing of multiple ratings through FHE operations.

\section{Conclusion}\label{sec:VI}
In this work, we introduce a novel approach for end-to-end privacy-preserving recommendation systems: CSR matrix factorization utilizing fully homomorphic encryption. This innovative architecture effectively addresses sparsity challenges inherent in large recommendation datasets while minimizing communication costs with high accuracy. 
Notably, our proposed CSR-based matrix factorization methodologies successfully resolve sparsity-related challenges within FHE domain, representing significant progress in privacy-preserving recommendation technology.
\par While our CSR-based matrix factorization efficiently supports privacy-preserving recommendation on encrypted sparse rating matrices, it lacks diagonal packing optimizations demonstrated in Garimella et al.\cite{b41} and multithreaded acceleration techniques used in Ferguson et al.'s \cite{b42} for sparse matrix multiplication schemes for FHE in deep neural network (DNN) inference. In future work, our aim is to integrate batched SIMD-style diagonal packing and multithreaded execution to reduce computational latency. 

\section{Acknowledgment}
This project is partially supported by NSF \#2451744.



\begin{thebibliography}{00}
\bibitem{b1} Su, Xiaoyuan, and Taghi M. Khoshgoftaar. ``A survey of collaborative filtering techniques.'' Advances in artificial intelligence 2009.1 (2009): 421425.
\bibitem{b2} Ramakrishnan, Naren, et al. ``When being weak is brave: Privacy issues in recommender systems.'' Technical paper posted on the Computing Research Repository at http://xxx. lanl. gov/abs/cs. CG/0105028 (2001).
\bibitem{b3} Narayanan, Arvind, and Vitaly Shmatikov. ``Robust de-anonymization of large sparse datasets.'' 2008 IEEE Symposium on Security and Privacy (sp 2008). IEEE, 2008.
\bibitem{b4}Weinsberg, Udi, et al. ``BlurMe: Inferring and obfuscating user gender based on ratings.'' Proceedings of the sixth ACM conference on Recommender systems. 2012.
\bibitem{b5} Canny, John. ``Collaborative filtering with privacy.'' Proceedings 2002 IEEE symposium on security and privacy. IEEE, 2002.
\bibitem{b6} McSherry, Frank, and Ilya Mironov. ``Differentially private recommender systems: Building privacy into the netflix prize contenders.'' Proceedings of the 15th ACM SIGKDD international conference on Knowledge discovery and data mining. 2009.
\bibitem{b7} Nikolaenko, Valeria, et al. ``Privacy-preserving matrix factorization.'' Proceedings of the 2013 ACM SIGSAC conference on Computer \& communications security. 2013.
\bibitem{b8}Shen, Yilin, and Hongxia Jin. ``Privacy-preserving personalized recommendation: An instance-based approach via differential privacy.'' 2014 IEEE international conference on data mining. IEEE, 2014.
\bibitem{b9}Sweeney, Latanya. ``k-anonymity: A model for protecting privacy.'' International journal of uncertainty, fuzziness and knowledge-based systems 10.05 (2002): 557-570.
\bibitem{b10}Jumonji, Seiya, et al. ``Privacy-preserving collaborative filtering using fully homomorphic encryption.'' IEEE Transactions on Knowledge and Data Engineering 35.3 (2021): 2961-2974.
\bibitem{b11}Kim, Sungwook, et al. ``Efficient privacy-preserving matrix factorization via fully homomorphic encryption.'' Proceedings of the 11th ACM on Asia conference on computer and communications security. 2016.
\bibitem{b12}Benaissa, Ayoub, et al. ``Tenseal: A library for encrypted tensor operations using homomorphic encryption.'' arXiv preprint arXiv:2104.03152 (2021).
\bibitem{b13}Cheon, Jung Hee, et al. ``Homomorphic encryption for arithmetic of approximate numbers.'' Advances in cryptology–ASIACRYPT 2017: 23rd international conference on the theory and applications of cryptology and information security, Hong kong, China, December 3-7, 2017, proceedings, part i 23. Springer International Publishing, 2017.
\bibitem{b14} Grouplens, Grouplens Moivelens 100k dataset [Online]. Available: https://files.grouplens.org/datasets/movielens/ml-100k
\bibitem{b15}Gemulla, Rainer, et al. ``Large-scale matrix factorization with distributed stochastic gradient descent.'' Proceedings of the 17th ACM SIGKDD international conference on Knowledge discovery and data mining. 2011.
\bibitem{b16}Gentry, Craig. ``Fully homomorphic encryption using ideal lattices.'' Proceedings of the forty-first annual ACM symposium on Theory of computing. 2009.
\bibitem{b17}Chevallier-Mames, Benoît, Pascal Paillier, and David Pointcheval. ``Encoding-free ElGamal encryption without random oracles.'' Public Key Cryptography-PKC 2006: 9th International Conference on Theory and Practice in Public-Key Cryptography, New York, NY, USA, April 24-26, 2006. Proceedings 9. Springer Berlin Heidelberg, 2006.
\bibitem{b18}Yao, Andrew C. ``Protocols for secure computations.'' 23rd annual symposium on foundations of computer science (sfcs 1982). IEEE, 1982.
\bibitem{b19}Huang, Yan, et al. ``Faster secure {Two-Party} computation using garbled circuits.'' 20th USENIX Security Symposium (USENIX Security 11). 2011.
\bibitem{b20}Nayak, Kartik, et al. ``Graphsc: Parallel secure computation made easy.'' 2015 IEEE symposium on security and privacy. IEEE, 2015.
\bibitem{b21}Rosinosky, Guillaume, et al. ``PProx: efficient privacy for recommendation-as-a-service.'' Proceedings of the 22nd International Middleware Conference. 2021.
\bibitem{b22}Yang, Liu, et al. ``Federated recommendation systems.'' Federated Learning: Privacy and Incentive. Cham: Springer International Publishing, 2020. 225-239.
\bibitem{b23}Kim, Jinsu, et al. ``Efficient privacy-preserving matrix factorization for recommendation via fully homomorphic encryption.'' ACM Transactions on Privacy and Security (TOPS) 21.4 (2018): 1-30.
\bibitem{b24}Smart, Nigel P., and Frederik Vercauteren. ``Fully homomorphic SIMD operations.'' Designs, codes and cryptography 71 (2014): 57-81.
\bibitem{b25}Mohammed, Noman, et al. ``Differentially private data release for data mining.'' Proceedings of the 17th ACM SIGKDD international conference on Knowledge discovery and data mining. 2011.
\bibitem{b26}Shen, Yilin, and Hongxia Jin. ``Privacy-preserving personalized recommendation: An instance-based approach via differential privacy.'' 2014 IEEE international conference on data mining. IEEE, 2014.
\bibitem{b27}Martins, Paulo, Leonel Sousa, and Artur Mariano. ``A survey on fully homomorphic encryption: An engineering perspective.'' ACM Computing Surveys (CSUR) 50.6 (2017): 1-33.
\bibitem{b28}Armknecht, Frederik, et al. ``A guide to fully homomorphic encryption.'' Cryptology ePrint Archive (2015).
\bibitem{b29}Marcolla, Chiara, et al. ``Survey on fully homomorphic encryption, theory, and applications.'' Proceedings of the IEEE 110.10 (2022): 1572-1609.
\bibitem{b30}Wu, Wenbin, et al. ``An Enhanced Latent Factor Recommendation Approach for Sparse Datasets of E-Commerce Platforms.'' Systems 13.5 (2025): 372.
\bibitem{b31}Brakerski, Zvika, Craig Gentry, and Vinod Vaikuntanathan. ``(Leveled) fully homomorphic encryption without bootstrapping.'' ACM Transactions on Computation Theory (TOCT) 6.3 (2014): 1-36.
\bibitem{b32}Fan, Junfeng, and Frederik Vercauteren. ``Somewhat practical fully homomorphic encryption.'' Cryptology ePrint Archive (2012).
\bibitem{b33}Chillotti, Ilaria, et al. ``TFHE: fast fully homomorphic encryption over the torus.'' Journal of Cryptology 33.1 (2020): 34-91.
\bibitem{b34}Veugen, Thijs, et al. ``A framework for secure computations with two non-colluding servers and multiple clients, applied to recommendations.'' IEEE Transactions on Information Forensics and Security 10.3 (2014): 445-457.
\bibitem{b35}``Microsoft SEAL (release 4.1)'', Jan. 2023. https://github.com/microsoft/SEAL (accessed May 25, 2025).
\bibitem{b36}Lak, Parisa, Bora Caglayan, and Ayse Basar Bener. ``The impact of basic matrix factorization refinements on recommendation accuracy.'' 2014 IEEE/ACM International Symposium on Big Data Computing. IEEE, 2014.
\bibitem{ufc}Zhou, Minxuan, et al. ``UFC: A Unified Accelerator for Fully Homomorphic Encryption.'' 2024 57th IEEE/ACM International Symposium on Microarchitecture (MICRO). IEEE, 2024.
\bibitem{fhemem}Zhou, Minxuan, et al. ``FHEmem: A processing in-memory accelerator for fully homomorphic encryption.'' IEEE Transactions on Emerging Topics in Computing (2025).
\bibitem{f1}Samardzic, Nikola, et al. ``F1: A fast and programmable accelerator for fully homomorphic encryption.'' MICRO-54: 54th Annual IEEE/ACM International Symposium on Microarchitecture. 2021.
\bibitem{sharp}Kim, Jongmin, et al. ``SHARP: A short-word hierarchical accelerator for robust and practical fully homomorphic encryption.'' Proceedings of the 50th Annual International Symposium on Computer Architecture. 2023.
\bibitem{b41}Garimella, Karthik, et al. ``HE-LRM: Encrypted Deep Learning Recommendation Models using Fully Homomorphic Encryption." arXiv preprint arXiv:2506.18150 (2025).
\bibitem{b42}Ferguson, Aidan, et al. ``Exploiting Unstructured Sparsity in Fully Homomorphic Encrypted DNNs." Proceedings of the 5th Workshop on Machine Learning and Systems. 2025.


\end{thebibliography}
\end{document}